\documentclass[fleqn,usenatbib]{mnras}

\usepackage{newtxtext,newtxmath}

\usepackage[T1]{fontenc}

\DeclareRobustCommand{\VAN}[3]{#2}
\let\VANthebibliography\thebibliography
\def\thebibliography{\DeclareRobustCommand{\VAN}[3]{##3}\VANthebibliography}

\usepackage{graphicx}	
\usepackage{amsmath}	
\usepackage{comment}
\usepackage{subfigure}
\usepackage{soul}

\newcommand{\parensuper}[2]{{#1}^{(#2)}}
\newcommand{\PlnkOpac}{\chi_\mathrm{p}}
\newcommand{\ixpe}{\emph{IXPE}}

\title[Signatures of particle bombardment]{X-ray polarisation signatures in bombarded magnetar atmospheres}

\author[R. M. E. Kelly et al.]{
Ruth M. E. Kelly,$^{1}$\thanks{E-mail: ruth.kelly.22@ucl.ac.uk}
Denis Gonz\'alez-Caniulef,$^{2}$
Silvia Zane,$^{1}$
Roberto Turolla$^{1,3}$
and Roberto Taverna$^{3}$
\\
$^{1}$Mullard Space Science Laboratory, University College London, Holmbury St Mary, Dorking, Surrey RH5 6NT\\
$^{2}$Institut de Recherche en Astrophysique et Planétologie, 9 avenue du Colonel Roche, BP 44346 31028, Toulouse CEDEX 4, France\\
$^{3}$Universit\`a di Padova, Dipartimento di Fisica e Astronomia, via Marzolo 8, I-35131 Padova, Italy
}

\date{Accepted XXX. Received YYY; in original form ZZZ}

\pubyear{2024}

\begin{document}
\label{firstpage}
\pagerange{\pageref{firstpage}--\pageref{lastpage}}
\maketitle

\begin{abstract}
Magnetars are neutron stars that host huge, complex magnetic fields which require supporting currents to flow along the closed field lines. This makes magnetar atmospheres different from those of passively cooling neutron stars because of the heat deposited by backflowing charges impinging on the star surface layers. This particle bombardment is expected to imprint the spectral and, even more, the
polarisation properties of the emitted thermal radiation. 
We present solutions for the radiative transfer problem for bombarded plane-parallel atmospheres in the high magnetic field regime. The temperature profile is assumed a priori, and selected in such a way to reflect the varying rate of energy deposition in the slab (from the impinging currents and/or from the cooling crust). 
We find that thermal X--ray emission powered entirely by the energy released in the atmosphere by the magnetospheric back--bombardment is linearly polarised and X-mode dominated,  but its polarisation degree is significantly reduced (down to $10\%$--$50\%$) when compared with that expected from a standard atmosphere heated only from the cooling crust below. 
By increasing the fraction of heat flowing in from the crust the polarisation degree of the emergent radiation increases, first at higher energies ($\sim 10\ \mathrm{keV}$) and then in the entire soft X-ray band.
We use our models inside a ray-tracing code to derive the expected emission properties as measured by a distant observer and compare our results with recent \ixpe\ observations of magnetar sources.
\end{abstract}




\section{Introduction}
\label{sec:intro}
Powered by their own magnetic energy, soft $\gamma$-repeaters (SGRs) and anomalous X-ray pulsars (AXPs) are two groups of isolated neutron stars which together make up the class known as magnetars \citep{duncan_formation_1992, thompson_neutron_1993}.
They are typically characterised by persistent X-ray luminosities $L\approx 10^{31}$--$10^{36}\,\mathrm{erg\,s}^{-1}$, spin periods $P\sim 1$--$12\,\mathrm{s}$ \cite[with the exception of the transient source 3XMM J185246.6+003317, that has been proposed to have a period of $23\,\mathrm{s}$, see][]{hambaryan_3xmm_2015} and exhibit period derivatives $\dot{P}\sim 10^{-13}$--$10^{-10}\,\mathrm{s\,s}^{-1}$, from which ultra strong magnetic fields $B\sim 10^{13}$--$10^{15}\,\mathrm{G}$ can be inferred.

Magnetars are expected to have internal magnetic fields more complex than a pure dipole, with  both a poloidal  and a toroidal component. The stresses exerted on the crust by the latter induce the formation of local twists in the external magnetic field, which becomes non-potential and needs to be supported by 
currents flowing along the closed field lines. Photons emitted from the cooling surface undergo repeated resonant Compton scatterings (RCSs) with the moving particles, producing a thermal+power-law spectrum in agreement with observations \cite[the RCS paradigm;][see also \citealt{turolla_magnetars_2015}]{thompson_electrodynamics_2002}. 

Due to the strength of the magnetic field, 
magnetar emission is expected to be linearly polarised in two normal modes, referred to as the ordinary (O), with electric field vector oscillating in the plane of the propagation direction and the local magnetic field, and the extraordinary (X) one, with the electric vector oscillating perpendicular to the same plane.
The large difference in the opacities, which are much lower for the X-mode, is at the basis of the expected large polarisation of radiation coming from a strongly magnetised plasma \cite[see e.g.][]{harding_physics_2006, potekhin_atmospheres_2017, gonzalez_caniulef_polarized_2016}.
Additionally, vacuum also contributes to shape the optical properties of the medium. This results in the occurrence of vacuum birefringence and the appearance of the vacuum resonance, where the polarisation modes may switch \citep{adler_photon_1971, lai_transfer_2003, ho_ii_2003}. Both these effects strongly influence the polarisation of radiation as measured at infinity. 

Imaging X-ray Polarimetry Explorer \cite[\ixpe;][]{weisskopf_imaging_2022}, a joint NASA-ASI space mission and the first telescope devoted to a systematic study of the sky in polarised X-rays, has observed four magnetars to date; the AXPs 4U 0142+61 \citep{taverna_polarized_2022}, 1RXS J170849.0$-$400910 \cite[hereafter 1RXS J1708 for short,][]{zane_strong_2023} and 1E 2259+586 \citep{heyl_detection_2024}, and the SGR 1806$-$20 \citep{turolla_ixpe_2023}. Polarisation has been detected for all the sources, albeit to different levels of confidence \cite[see also][for a review]{taverna_x-ray_2024}.

Both 4U 0142+61 and 1RXS J1708 show polarisation properties which are highly energy dependent. 1RXS J1708 was found to have a constant polarisation angle while the polarisation degree increases from $\approx20 \%$ at $2$--$3\ \mathrm{keV}$ to $\approx 80\%$ at $6$--$8\ \mathrm{keV}$ \cite[]{zane_strong_2023}. On the other hand, 4U 0142+61 exhibits a polarisation degree of $\approx 15\%$ at $2$--$4\ \mathrm{keV}$ and $\approx 35\%$ at $6$--$8\ \mathrm{keV}$, dropping to zero at $4$--$5\ \mathrm{keV}$ where the polarisation angle swings by $90^{\circ}$, indicating a switch of the dominant polarisation mode from low to high energies \cite[]{taverna_polarized_2022}. Only a marginally significant polarisation degree of $31.6 \pm 10.5 \%$ was detected in the $4$--$5\ \mathrm{keV}$ range for SGR 1806$-$20, with $3\sigma$ upper limits of $24 \%$ and $55\%$ in the $2$--$4$ and $5$--$8$ keV bands, respectively \cite[]{turolla_ixpe_2023}. 1E 2259+586 was found to have a mild, phase-dependent polarisation degree, ranging from 
$\approx 0$ to $\approx 25 \%$ \cite[][]{heyl_detection_2024}.
 
The polarisation properties from both 4U 0142+61 and SGR 1806$-$20 can be explained by the reprocessing of thermal radiation from a condensed surface by RCS in the star twisted magnetosphere \cite[]{taverna_x-ray_2020, taverna_polarized_2022, turolla_ixpe_2023}. 
The polarisation signature of 1RXS J1708 is, instead, compatible with thermal emission coming from two regions of the surface, one covered by a standard atmosphere, the other in a magnetically-condensed state  \cite[]{zane_strong_2023}. An alternative model has been proposed by \cite{lai_ixpe_2023}, according to which the polarisation pattern of 4U 0142+61 and 1RXS J1708 could be due to partial mode conversion at the vacuum resonance in the star atmosphere. A recent, more comprehensive study of atmospheres with partial mode conversion, however, questions this result \cite[][]{kelly_x-ray_2024}. Finally, a baryon-loaded magnetic loop \cite[similar to that proposed for SGR 0418+5729 by][]{tiengo_variable_2013} can explain the phase-dependent absorption line and polarisation properties of  1E 2259+586 \cite[]{heyl_detection_2024}. 

The suggestion that 
the low-energy spectral components observed by \ixpe\  originate in the solid crust was put forward mainly to explain two features: a) the relatively modest polarisation, and b) the fact that they appear to be polarised either in the O mode (as in 4U 0142+61 and 1E 2259+586) or in the X mode (as in 1RXS J1708). These two facts are difficult to explain with standard atmospheric emission from a passive cooler. However, it is possible that atmospheric models computed under different assumptions provide an alternative, viable  interpretation to the solid surface scenario.  For instance,
atmospheres around active magnetars are expected to be at variance with those of other (strongly) magnetised NSs, inasmuch the magnetospheric currents (needed to sustain the twisted field) impact on the star outer layers, depositing heat, a phenomenon known in literature as ``particle bombardment''. 
It has been proposed \cite[see e.g.][]{gonzalez-caniulef_atmosphere_2019, 
mushtukov_spectrum_2021, taverna_polarized_2022, taverna_x-ray_2024} that the extra heat deposited in the external surface layers may produce a) a substantial reduction in the polarisation of the emergent radiation, and b) inner atmospheric layers characterised by an ``inverted temperature profile'' (with the temperature decreasing at larger depths), which, in turn, can give rise to O-dominated emergent radiation (see \S~2). A self-consistent modelling of magnetar atmospheres, accounting for the bombardment effect, is therefore warranted to investigate this possibility and confront spectro-polarimetric data. 

An initial investigation into this effect  has been carried out by \cite{gonzalez-caniulef_atmosphere_2019} who studied the problem of the thermal structure of a gray, bombarded atmosphere. In this work we expand upon these first results, exploring to what extent particle bombardment influences the frequency dependent properties of thermal radiation emitted by the atmosphere. 

The paper is laid out as follows. In section \ref{sec:theory} we review the basic characteristics of the expected  atmospheric temperature profile under different assumptions, the physics of heat deposition by particle bombardment, the atmospheric model calculations and the assumptions made throughout our investigation. We present our numerical results in section \ref{sec:results} and compare our findings with observed polarisation signatures in section \ref{sec:application}. Finally, discussion and conclusions are presented in section \ref{sec:discussion}.

\section{Theoretical Background}
\label{sec:theory}
\subsection{Atmospheric temperature profile}
\label{sec:temp-prof}

The aim of this paper is to study the propagation of radiation in a magnetised NS atmosphere accounting for different temperature profiles,
with a focus  
on how the temperature gradient affects the polarisation signal. We do not self-consistently solve the energy balance and radiation field, since this  would require the development of a new numerical code, which  
is beyond the scope of the present work. The temperature run, instead, is specified a priori, starting from the  results obtained in previous investigations on both passive and bombarded atmospheres. For the sake of completeness, these are reviewed below. 

A passively cooling neutron star, for which only the release of internal heat is responsible for the thermal emission, can be assumed to be covered by an atmosphere that is in local thermodynamic, radiative and hydrostatic equilibrium. Under these assumptions, atmospheric models have been computed by several authors
\cite[see e.g.][]{hernquist_thermal_1985, romani_model_1987, pavlov_multiwavelength_1996, ho_atmospheres_2001}. 
The energy balance equation can be expressed in the form
\begin{equation}
    \PlnkOpac\biggl[\frac{aT^4}{2}-\frac{\parensuper{\PlnkOpac}{1} \parensuper{U}{1}}{\PlnkOpac}-\frac{\parensuper{\PlnkOpac}{2} \parensuper{U}{2}}{\PlnkOpac}\biggl] = 0\,,
\end{equation}
where, $\chi_\mathrm p^\mathrm{(j)}$ and $U^\mathrm{(j)}$ are the Planck mean opacity and the energy density, respectively, referred to the normal mode $j$ (here $j=1$ is the extraordinary, X, and $j=2$ is the ordinary, O, one), while $\chi_\mathrm p$ is the total Planck mean opacity.
The basic feature of the resulting temperature profile is that it increases monotonically with increasing depth.
Due to different frequencies decoupling at different temperatures, this profile produces a typical hardening in the spectrum.
Owing to the lower opacity, the X-mode decouples at higher temperatures and is therefore expected to dominate the emergent radiation signal \cite[]{pavlov_model_1994, harding_physics_2006, potekhin_atmospheres_2014}.

As stated above, the situation can be quite different in the case of  magnetars, since returning currents may provide another source of heat, which is deposited as the back-flowing charges hit the atmosphere. 
This effect has been discussed by \cite{gonzalez-caniulef_atmosphere_2019} in the low-density plasma regime. Specifically, these authors computed frequency-integrated (gray) numerical models under the assumption that the entire observed luminosity is due to the release of the heat deposited in the external surface layers. They also assumed that  heat is deposited uniformly through a layer that, from the top of the atmosphere, extends down to a column density $y_0$, corresponding to the ``characteristic stopping length'' of the impinging particles; $y_0$ is assumed to be a constant and is taken as a model parameter. In strict analogy to the approach used in the literature when investigating stationary, accreting atmospheres \cite[see e.g.][]{turolla_spherical_1994, zampieri_x-ray_1995, zane_hot_1998},  the energy balance equation is modified into 
\begin{equation}\label{en-bal-bomb}
    \frac{\PlnkOpac}{\chi_\mathrm{sc}}\biggl[\frac{aT^4}{2}-\frac{\parensuper{\PlnkOpac}{1} \parensuper{U}{1}}{\PlnkOpac}-\frac{\parensuper{\PlnkOpac}{2} \parensuper{U}{2}}{\PlnkOpac}\biggl] + (\Gamma - \Lambda)_\mathrm C = \frac{W_\mathrm H}{c\chi_\mathrm{sc}}\,,
\end{equation}
where $(\Gamma-\Lambda)_\mathrm{C}$ is the Compton heating-cooling term, the inclusion of which is necessary to ensure the energy balance in the external layers. 
In the above expression $\chi_\mathrm{sc}$ is the mean scattering opacity and $W_\mathrm H$ can be approximated by
\begin{equation}
        W_\mathrm H =\left\{\begin{array}{lll} 
        \dfrac{L_{\infty}}{4\pi R^2y_0} &\; \; \; & y < y_0 \\
        \ & \ &\ \\
        0 &\; \; \;& y \geq y_0
        \end{array}\right.\, , 
\end{equation}
where $L_\infty$ is the luminosity observed at infinity 
and $R$ is the star radius.
The local luminosity at the atmosphere surface is
\begin{equation}
        L = L_\infty \biggl(1-\frac{2GM}{Rc^2}\biggl)^{-1}\,,
\end{equation}
where $M$ is the neutron star mass.
The main problem is then the calculation of the characteristic stopping length.
\cite{gonzalez-caniulef_atmosphere_2019} describes in detail the chain of processes that are expected to contribute to the particle deceleration. Briefly summarised, impinging particles travel along the magnetic field lines, scatter and can release a photon. The process is then repeated through a new scattering until complete deceleration of the primary particle. The released photons can trigger an electron-positron avalanche and the energetic secondary particles that are produced heat the layers as they propagate further through the atmosphere.

\begin{figure}
    \centering
    \subfigure{\includegraphics[width=\columnwidth]{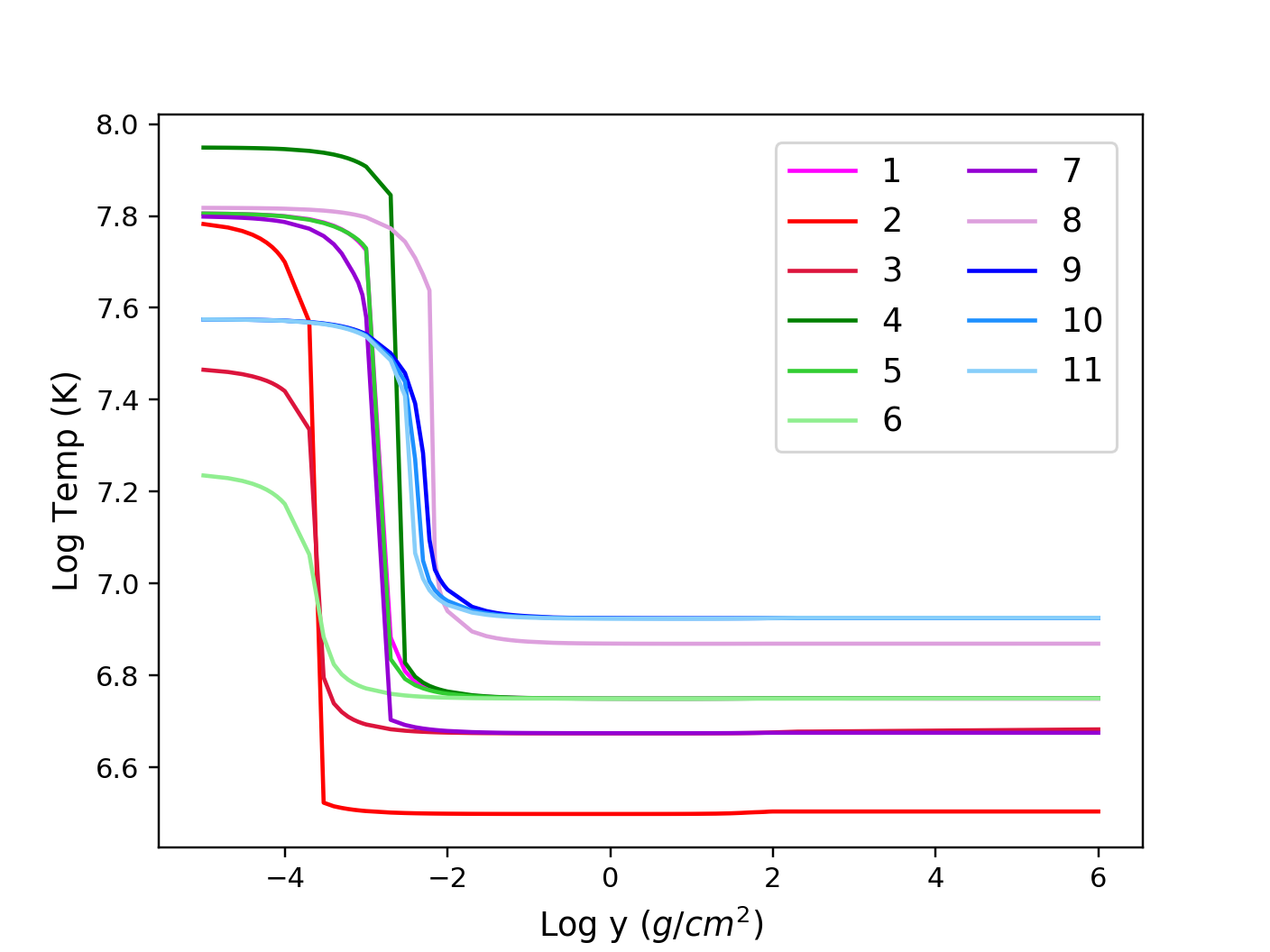}}
    \caption{Temperature as a function of column density for atmospheres experiencing particle bombardment for different magnetic field strengths $B$, luminosities $L_\infty$ and stopping column densities $y_0$. The parameters values for the different models are reported in Table \ref{tab:modelvalues}.}
    \label{fig:DenisTemp}
\end{figure}

These processes, alongside Compton drag, affect the value of the ``stopping length''.  It is worth noting that what we refer to as the ``stopping length'' is not merely the stopping length of the impacting particle; instead it refers to the length over which the heat is deposited in the atmosphere due to the chain of processes described above.
These  mechanisms have been discussed in detail 
by \cite{gonzalez-caniulef_atmosphere_2019}, who produced numerical simulations and concluded that, for impinging  particles with Lorentz factor $\gamma = 10^{3}$, the  characteristic stopping column density is in the range $y_{0} \approx 65$--$500\ \mathrm{g\, cm}^{-2}$.
Figure \ref{fig:DenisTemp} shows the temperature profile of atmospheres experiencing particle bombardment as a function of the column density, as computed by \cite{gonzalez-caniulef_atmosphere_2019}; the model parameters are listed in Table \ref{tab:modelvalues}.

A hot layer is formed at the surface of the atmosphere where free-free cooling is not sufficient to radiate the heat released by particle bombardment and  thermal equilibrium is instead maintained by Compton cooling. This hot external layer extends until the density is large enough for free-free cooling to kick in \cite[]{zane_magnetized_2000, gonzalez-caniulef_atmosphere_2019}. It is worth noting that the extension of this layer does not correspond to the stopping column density $y_0$.

As it can be seen,  varying $B$ results in no significant change to the temperature profile (see models 9, 10 and 11 in Figure \ref{fig:DenisTemp}). On the other hand, by increasing $L_\infty$ the hot external layer (where the heat deposited in the bombardment process is dissipated by Compton cooling) extends deeper into the atmosphere, producing higher temperatures (models 1 and 2 in Figure \ref{fig:DenisTemp}). Also, the external layer becomes hotter and wider by decreasing $y_0$ (see again Figure \ref{fig:DenisTemp}, models 5 and 6). 
The most striking feature, common to all cases, is however the presence of a nearly flat temperature profile in an extended zone of the atmosphere.

\begin{table}
\centering
    \caption{Parameter values for the models displayed in Figure \ref{fig:DenisTemp}.}
    \begin{tabular}{l c c c }
        \hline
        Model & $B$ ($\mathrm{G}$) & $L_\infty$ ($\mathrm{erg\,s}^{-1}$) & $y_0$ ($\mathrm{g\,cm}^{-2}$) \\
        \hline\hline
        1 & $3\times10^{14}$ & $10^{36}$ & $100$ \\
        \hline
        2 & $3\times10^{14}$ & $10^{35}$ & $100$ \\
        \hline
        3 & $4\times10^{14}$ & $10^{35}$ & $200$ \\
        \hline
        4 & $4\times10^{14}$ & $10^{36}$ & $70$ \\
        \hline
        5 & $4\times10^{14}$ & $10^{36}$ & $100$ \\
        \hline
        6 & $4\times10^{14}$ & $10^{36}$ & $500$ \\
        \hline
        7 & $3\times10^{14}$ & $5\times10^{35}$ & $100$ \\
        \hline
        8 & $3\times10^{14}$ & $3\times10^{36}$ & $100$ \\
        \hline
        9 & $3\times10^{14}$ & $5\times10^{36}$ & $200$ \\
        \hline
        10 & $6\times10^{14}$ & $5\times10^{36}$ & $200$ \\
        \hline
        11 & $8\times10^{14}$ & $5\times10^{36}$ & $200$ \\
        \hline\hline
    \end{tabular}
    \label{tab:modelvalues}
\end{table}

\begin{figure}
    \centering
    {\includegraphics[width=\columnwidth]{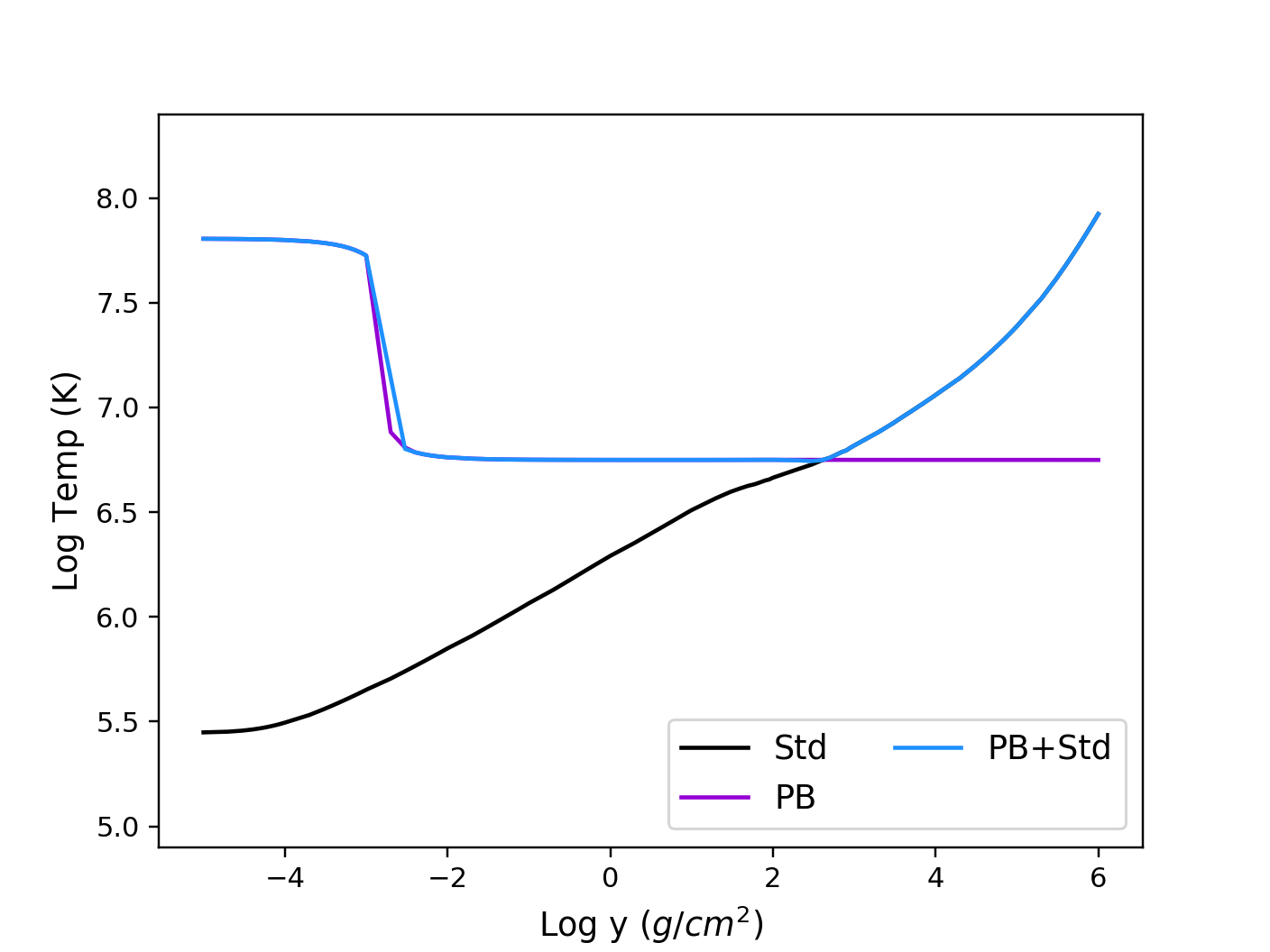}}
    \caption{Temperature as a function of atmospheric column density for a bombarded atmosphere (purple curve) and a standard cooling atmosphere (black curve). The cyan line shows a tentative profile for a bombarded atmosphere heated also from below. Here, $B=3\times10^{14}\ \mathrm{G}$, $L_\infty=10^{36}\ \mathrm{erg/s}$  and $y_0 = 100\ \mathrm{g/cm}^2$}
    \label{fig:extreme_temp}
\end{figure}

In  a more realistic scenario,  irradiation from the underlying crust can also contribute to the emitted luminosity, as is the case of a passive atmosphere. As a consequence, one qualitatively expects that at some depth inside the atmosphere 
the temperature will start raising inward in response to the heat flowing from the crust.  Figure~\ref{fig:extreme_temp} shows the temperature profile of a standard cooling atmosphere, self-consistently calculated with the numerical code by \cite{lloyd_model_2003} for  $B=3\times10^{14}\ \mathrm{G}$ and $L_\infty=10^{36}\ \mathrm{erg\,s}^{-1}$, together with a bombarded model with the same $B$ and $L_\infty$ (model 1 in Figure \ref{fig:DenisTemp}). A possible profile for a bombarded atmosphere heated also from below is shown by the cyan line. 

\begin{figure}
    \centering
    {\includegraphics[width=\columnwidth]{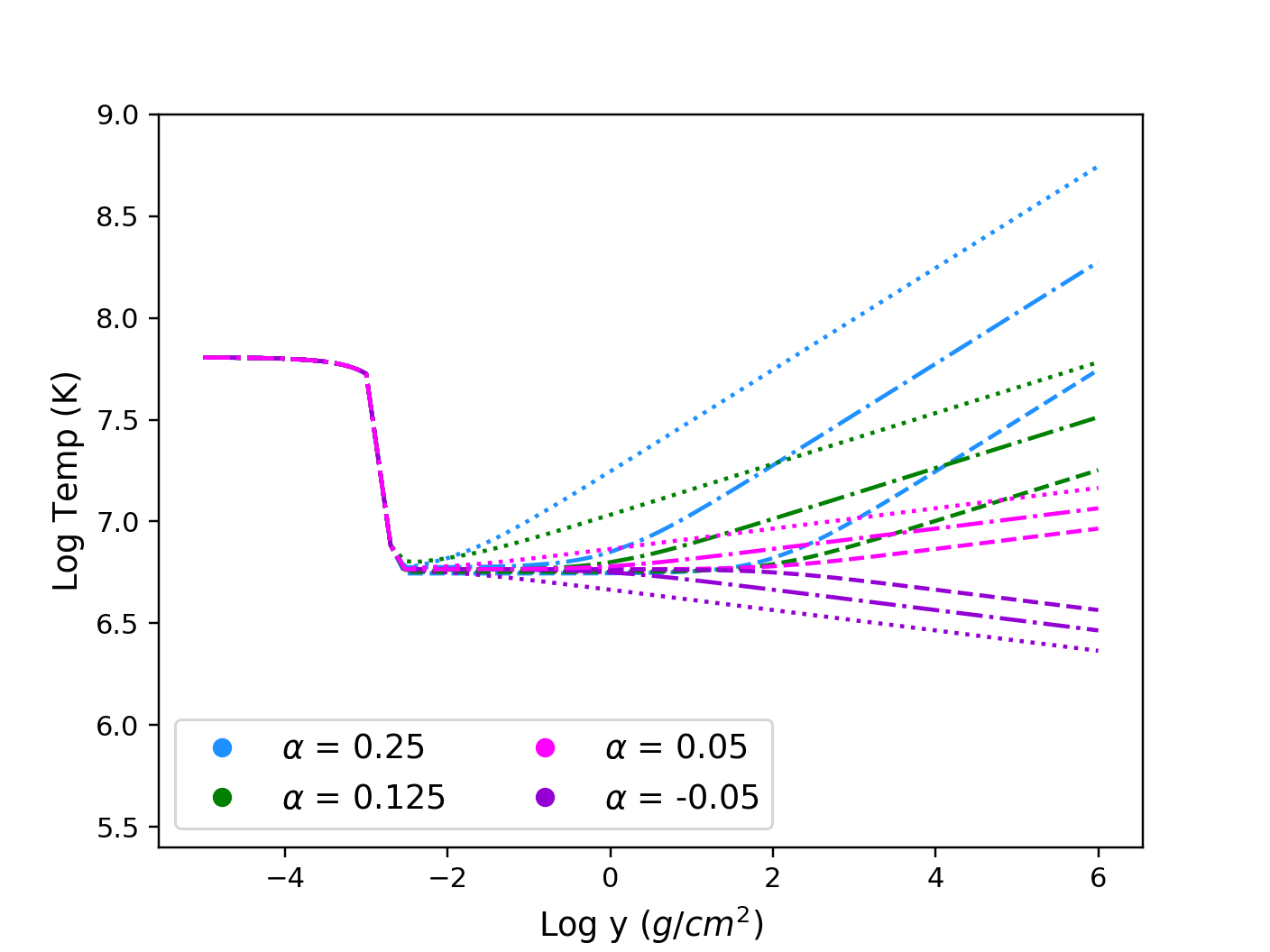}}
    \caption{Temperature profiles for different choices of the power-law slope $\alpha = 0.25, 0.125, 0.05, -0.05$ and matching depth $\log y_* = -2, 0, 2$ (dotted, dash-dotted, and dashed lines, respectively); here the external hot layer is taken from model 1 in Figure \ref{fig:DenisTemp}.
    }
    \label{fig:PL_temp}
\end{figure}

In a gray, unmagnetised, pure scattering atmosphere the temperature scales with depth as $\tau^{1/4}\sim y^{1/4}$. On the wake of these considerations, we adopt a simplified approach in which the inner temperature profile is a power-law  with index $\alpha$ ranging from $0.05$ to $0.25$ and starting at different atmospheric depths $y_*$. 
Clearly,  this is an oversimplification and is just useful for illustrative purposes.
It has been also speculated that  particle bombardment  can produce an ``inverted temperature profile'', with the temperature decreasing inward in some layers, and that this may be responsible for an excess of O-mode photons in the emitted radiation \cite[]{gonzalez-caniulef_atmosphere_2019, taverna_polarized_2022}. In order to investigate this, we also consider slightly negative slopes. The set of temperature models used in our calculations is shown in Figure \ref{fig:PL_temp}. The models are labelled in terms of `equivalent isotropic luminosity’ for visualisation purposes. This value is the model flux multiplied by $4\pi R^2$ and hence each curve can be re-scaled once a fraction of the emitting area is specified.

\subsection{Model regime}\label{sec:regime}

Our model relies on the standard twisted magnetosphere regime detailed in \cite{beloborodov_corona_2007, beloborodov_electron-positron_2013}. In this scenario, as is described in \cite{beloborodov_corona_2007}, the current needed to circulate around the magnetic field lines and sustain the twist cannot be carried by electrons/protons only. Would that be the case, the solution would be described by a ``relativistic double layer'' (the simple circuital analogue of a twisted, current-carrying bundle) in which the potential drop required to produce the conduction current is huge ($\Phi_\mathrm{PD}\sim 10^6$ GeV) and electrons are accelerated up to $\gamma \sim 10^9$. In this case, the twist would decay immediately. Instead, the reason a twist can be sustained is entirely due to the presence of pairs.

Surface thermal photons with energy $\epsilon$ resonantly scatter onto electrons when 

\begin{equation}
    \frac{\epsilon}{m_ec^2}\frac{\gamma}{1-\beta \cos{\theta}} = \frac{B}{B_Q},
\end{equation}
where $B_Q=4.413\times10^{13}\,\mathrm{G}$ is the quantum critical field, $\beta=v/c$ is the charge velocity (in units of the speed of light) along the magnetic field and $\theta$ is the angle between the photon direction and the magnetic field. In the inner magnetospheric region around a magnetar, where $B > 2B_Q$, $1$ keV photons (typically emitted by the neutron star surface) scatter onto $\gamma > 1000$ electrons. The scattered photons have energy $\epsilon’$ in the MeV range and initially propagate along the magnetic field. However, as soon as the photon trajectory deviates from the field line by a finite angle $\theta$, the condition

\begin{equation}
    \epsilon' > \frac{2m_ec^2}{\sin\theta}
\end{equation}
can be satisfied and the photons convert into pairs via 
\begin{equation}
    \gamma + B = e^+ + e^- + B.
\end{equation}

Pair production along the entire circuit efficiently screens the potential and the ``twist bundle'' current 
$j_B$ can then be conducted with $\Phi \ll \Phi_\mathrm{DL}$. As discussed in these papers \citep[see also][]{beloborodov_corona_2007, beloborodov_electron-positron_2013, torres_magnetar_2011, turolla_magnetars_2015}, a quasi-stationary state develops in which the particle energy has just the value required to ignite the pair cascade. For scattering onto soft X-ray photons ($\sim 1$ keV) this is $\gamma \sim 1000$.

As is shown by \cite{beloborodov_electron-positron_2013}, at larger distance from the star surface repeated scatterings can slow down the pairs, thus above a few neutron star radii the circuit contains mildly relativistic charges ($\gamma \sim 10$). It is nonetheless understood that before being able to create pairs in the inner region, electrons must reach $\gamma > 1000$, and so the parts of the twist bundle closer to the surface of the neutron star are populated by electrons (and positrons) with a Lorentz factor of that order. For our work we assume bombardment by a population of single velocity charges from close to the surface.

\subsection{Radiative transfer}\label{sec:radtra}

To calculate the emergent X-ray spectrum and polarisation degree we use the numerical code described in \cite{lloyd_model_2003}, with some modifications to deal with the case at hand. In particular, we turned off the self-consistent calculation of the temperature profile, so that the latter can be fed in input according to the prescription discussed in the previous section. The radiative transfer equations
\begin{equation}
    \mu\frac{dI_\nu^j(\boldsymbol k)}{\rho dz} = \chi_\nu^j(\boldsymbol k)I_\nu^j(\boldsymbol k) - \eta_\nu^j( \boldsymbol k)\,,
\end{equation}
are then solved for the two polarisation modes, assuming a geometrically thin plane-parallel atmosphere composed  of fully-ionised hydrogen.
Here, $\mu=\cos\theta_\mathrm{k}$, where $\theta_\mathrm{k}$ represents the angle the ray makes with the slab normal, the photon momentum $\boldsymbol k$ is described by $\mu$ and the azimuthal angle $\phi$, $I_\nu^j$ is the monochromatic intensity for mode $j$, and $\chi_\nu^j(\boldsymbol k)$ and $\eta_\nu^j( \boldsymbol k)$ are the monochromatic opacity and emissivity respectively.
Thomson scattering and bremsstrahlung emission/absorption are accounted for, while mode conversion at the vacuum resonance is neglected for simplicity. 
An iterative method is then used to calculate the scattering integrals \cite[see][for all details]{lloyd_model_2003}.

We use a grid of $100$ depth and $15$ $\mu$ points, and, at this stage, we compute models with an aligned magnetic field (so that the azimuthal angle  becomes unnecessary).
Our models are computed assuming a star mass $M=1.4\, M_\odot$ and radius $R= 10\, \mathrm{km}$, equating to a surface gravity $g=2.4\times10^{14}\, \mathrm{cm/s^2}$ \footnote{The original bombarded temperature profiles (see Section \ref{sec:temp-prof}) were computed using $M=1\, M_\odot$ as in \cite{gonzalez-caniulef_atmosphere_2019}. We checked that using $M=1\, M_\odot$  produces only marginal changes in our polarisation results ($\leq 2\%$)}.

Once the emergent intensity and its moments are computed, for the two modes of polarisation, we calculate the intrinsic polarisation degree of the emitted radiation,  defined as
\begin{equation}
    PD_\mathrm{em} = \frac{J^\mathrm X_\nu-J^\mathrm O_\nu}{J^\mathrm X_\nu+J^\mathrm O_\nu},
\end{equation}
where $J^\mathrm{X(O)}_\nu \equiv J^\mathrm{X(O)}_\nu\vert_{y=0}$ is the emergent mean monochromatic intensity of X (O) mode photons, such that positive $PD_\mathrm{em}$ indicates X-mode dominated emission.

\section{Results}
\label{sec:results}

In this section, we investigate how particle bombardment and different heating contributions from the crust affect the polarisation properties of the radiation emerging from a single slab of magnetar atmosphere.

\begin{figure}
    \centering
    \subfigure[]{\includegraphics[width=\columnwidth]{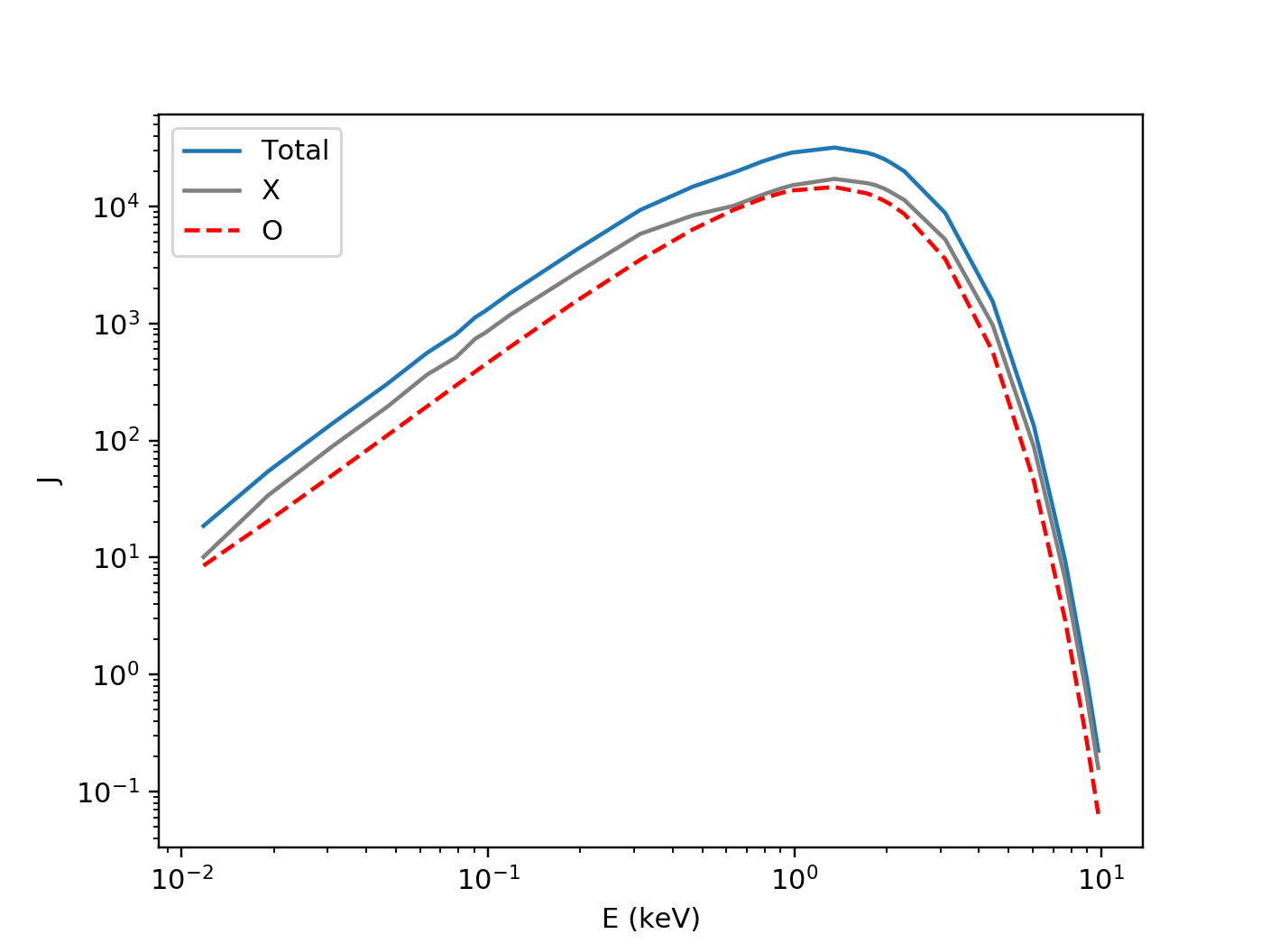}}
    \subfigure[]{\includegraphics[width=\columnwidth]{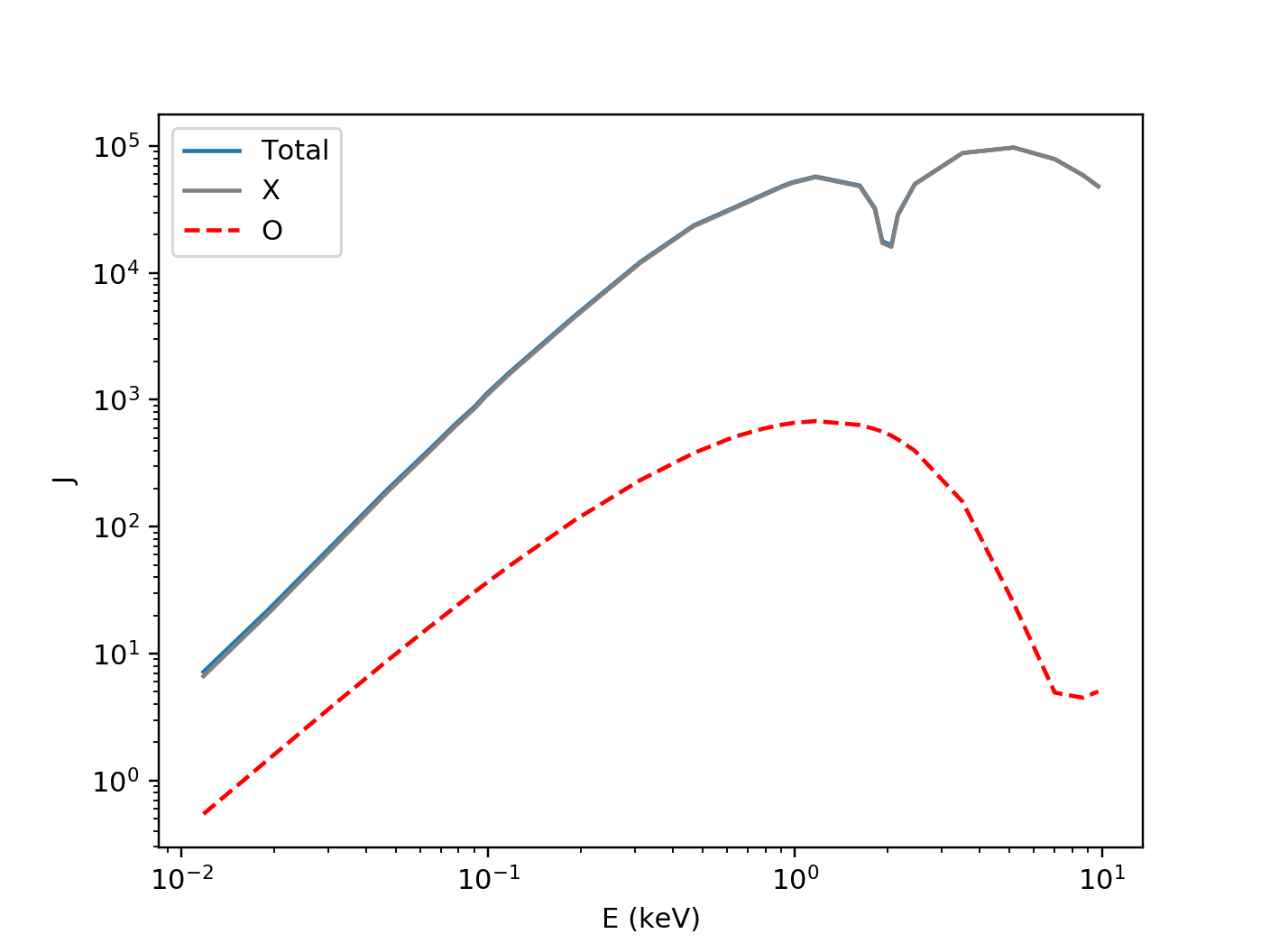}}
    \subfigure[]{\includegraphics[width=\columnwidth]{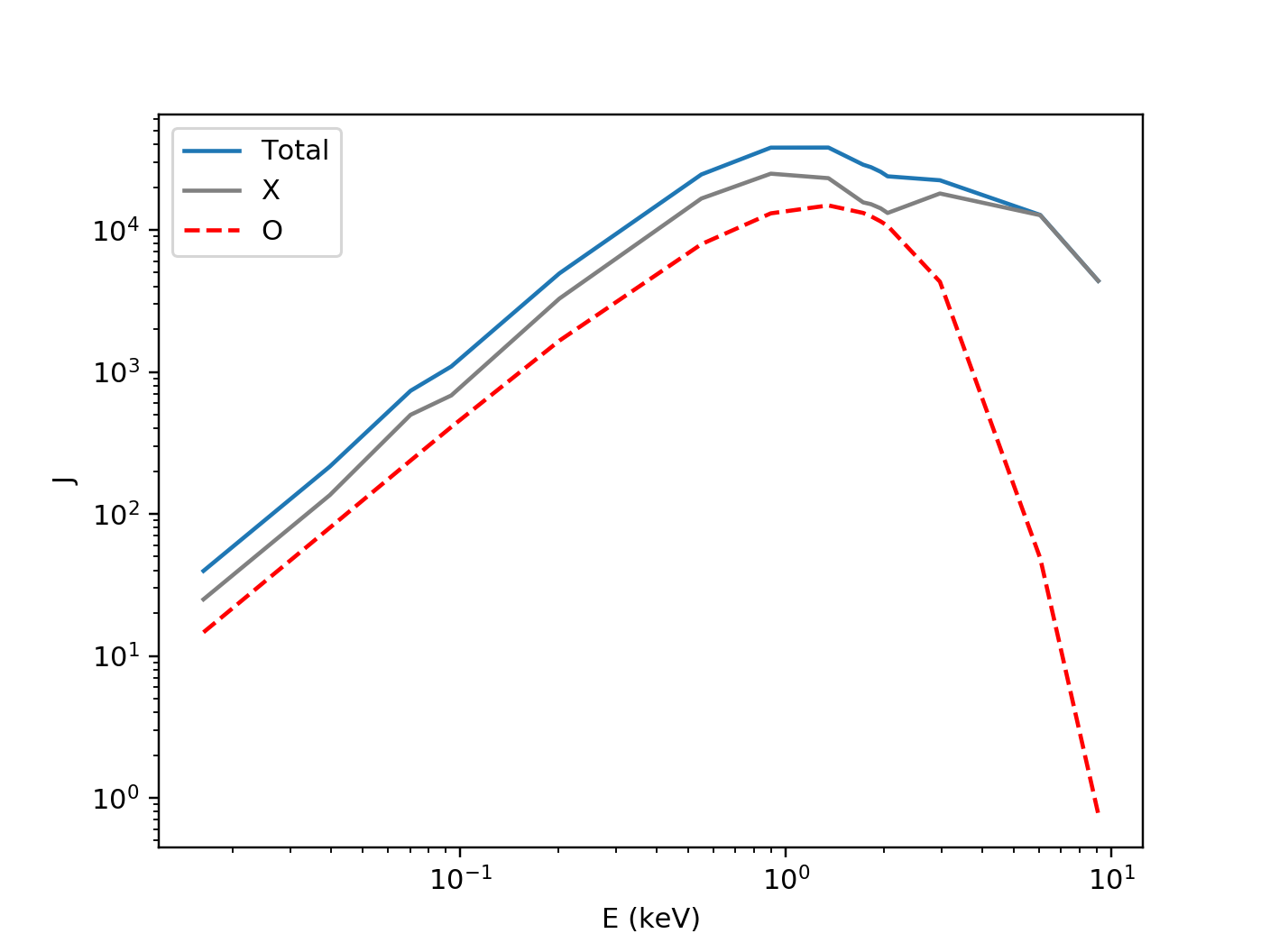}}
    \caption{Mean intensity (in $\mathrm{erg\ cm^{-2}\ s^{-1}\ keV^{-1} }$) spectrum of the emergent radiation as a function of the photon energy $E$ for the three cases described in Figure \ref{fig:extreme_temp}, where (a) is the bombarded profile, (b) is the standard cooling atmosphere and (c) is the bombarded atmosphere also heated from the crust.}
    \label{fig:spectra}
\end{figure}

Figure \ref{fig:spectra} shows the mean intensity spectra as a function of energy for the three temperature profiles presented in Figure \ref{fig:extreme_temp}. At variance with the case of a standard cooling atmosphere, panel (b), the spectrum from a bombarded model, panel (a), is more blackbody-like, the contributions of the two modes are closer and the proton cyclotron line (which for these parameters is at about $2\ \mathrm{keV}$) is not visible. These features are a direct consequence of the presence of an extended temperature plateau in the bombarded atmosphere (see again Figure \ref{fig:extreme_temp}). Since the X-mode opacity is much reduced with respect to that of the O-mode, X-mode photons decouple deeper in the layer and carry the imprint of the temperature there. If $T$ increases with depth (as in the standard cooling atmosphere), this results in a dominance of X-mode photons, which come from hotter regions, especially at higher energies where the free-free opacity drops. On the other hand, if the temperature is constant, X- and O-mode photons still come from different depths, but their spectra are much closer to each other (and closer to the same blackbody). The lack of the proton cyclotron feature in panel (a) can be explained likewise. The huge opacity at the resonance forces X-mode photons back to thermal equilibrium, so that  their spectrum approaches that of the O-mode. The appearance of a noticeable absorption feature is then related to how far apart the X- and O-mode continua are near the resonance. The spectrum for the combined temperature profile, panel (c), clearly exhibits intermediate properties among the two extreme cases discussed above. The cyclotron line is still present, albeit much reduced, and the X-mode dominates above $\sim 2\ \mathrm{keV}$ because photons of these energies come from the hotter regions, where $T$ increases inward.

\begin{figure}
    \centering
    {\includegraphics[width=\columnwidth]{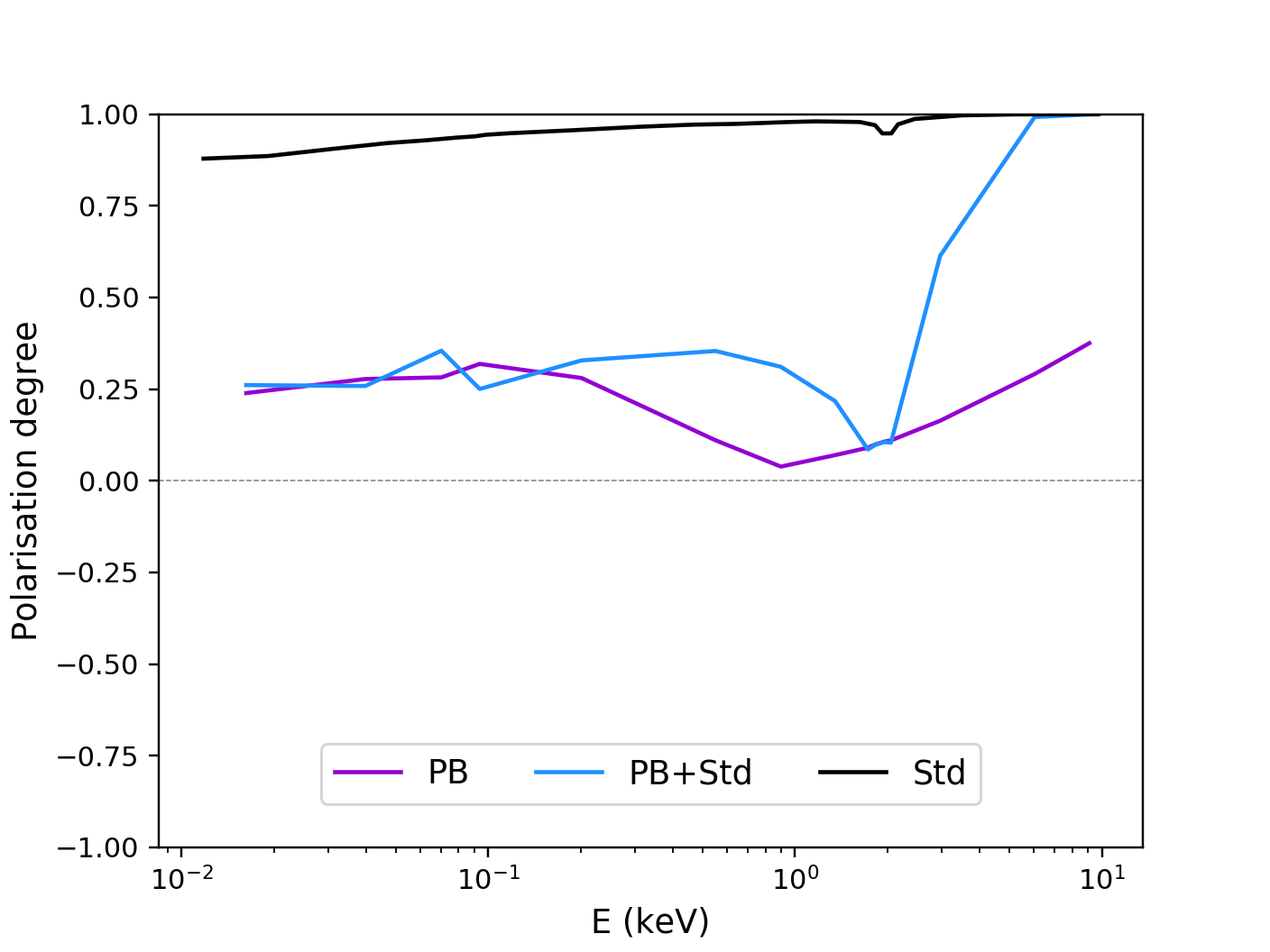}}
    \caption{Polarisation degree as a function of the photon energy $E$ for the models shown in Figure \ref{fig:extreme_temp}.}
    \label{fig:extreme}
\end{figure}

The energy-dependent polarisation degree for the three models discussed above is shown in Figure \ref{fig:extreme}. As expected from our previous considerations, emission from particle bombardment is polarised in the X-mode but the polarisation degree is  significantly lower ($\sim 25\%$) than in a standard cooling atmosphere (almost $100\%$).  
The model with  combined temperature profile  results in a signal with intermediate properties. The polarisation degree exhibits the characteristics of the bombarded model for energies below $\sim 0.2\ \mathrm{keV}$ after which it raises until it attains $\sim 100\%$ at $\sim 5\ \mathrm{keV}$. 
The drop in polarisation around $\sim 2\ \mathrm{keV}$ reflects the absorption of X-mode photons at the proton cyclotron resonance. This is much more pronounced than in the  passively cooling model because of the different X-to-O ratio (see also panels b and c in Figure \ref{fig:spectra}). 

We address next the polarisation properties of the bombarded atmosphere models  for all the temperature profiles presented in Figure \ref{fig:DenisTemp}, which cover different magnetic field strengths $B$, luminosities $L_\infty$ and stopping column densities $y_0$ (see also Table \ref{tab:modelvalues}). 
\begin{figure}
     \centering
    {\includegraphics[width=\columnwidth]{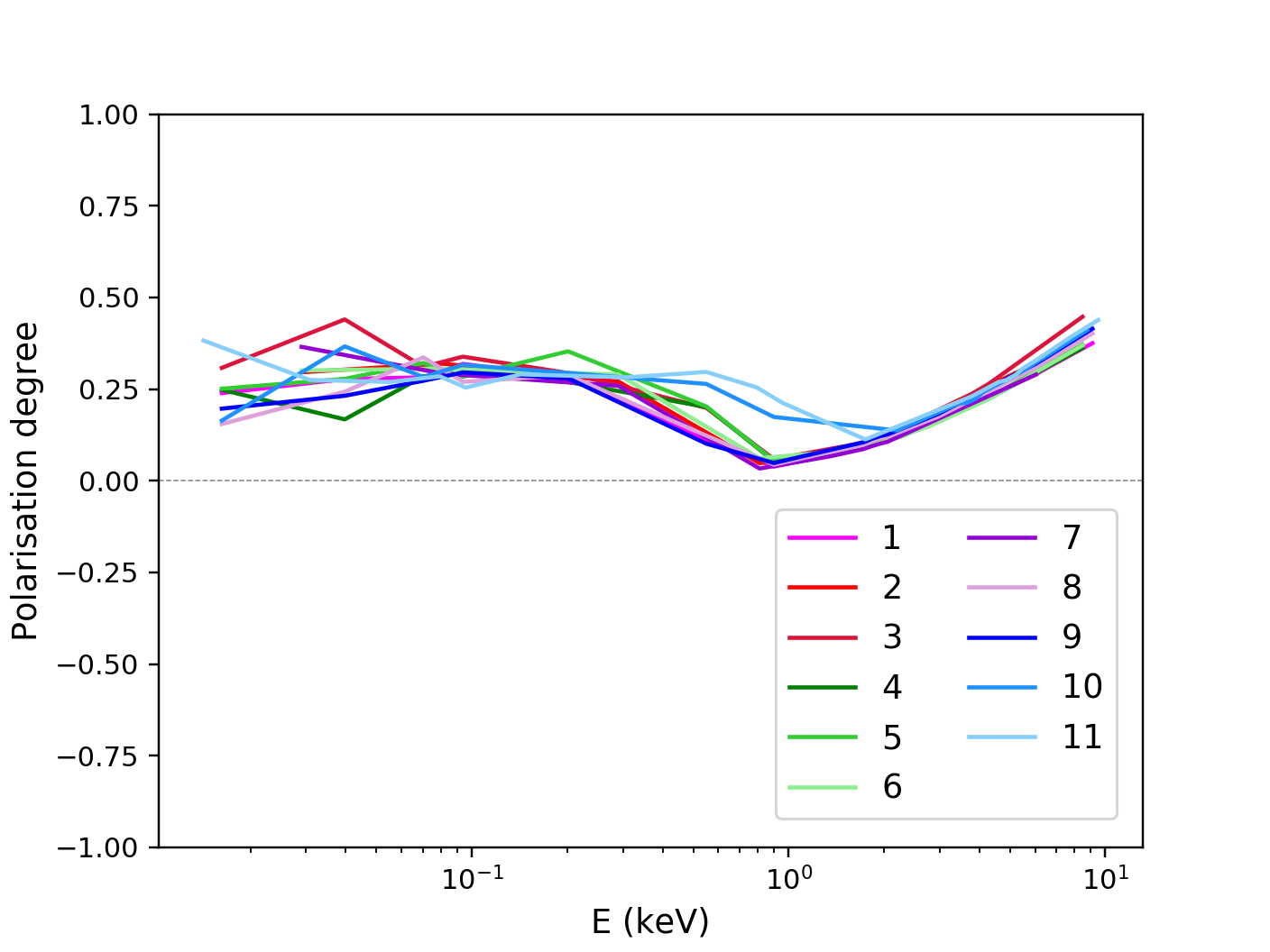}}
     \caption{Polarisation degree as a function of photon energy $E$ for atmospheres experiencing particle bombardment with temperature profiles shown in Figure \ref{fig:DenisTemp}.}
     \label{fig:DenisPol}
 \end{figure}
The corresponding polarisation degree, as a function of the photon energy, is shown in Figure \ref{fig:DenisPol}.
As it can be seen, all models essentially follow the qualitative behaviour exhibited by model 1 which is actually the one we presented in Figure \ref{fig:extreme}. The largest departures are for the most magnetised cases (models 10 and 11) which show a somewhat higher polarisation around $1\ \mathrm{keV}$. 

\begin{figure}
    \centering
    {\includegraphics[width=\columnwidth]{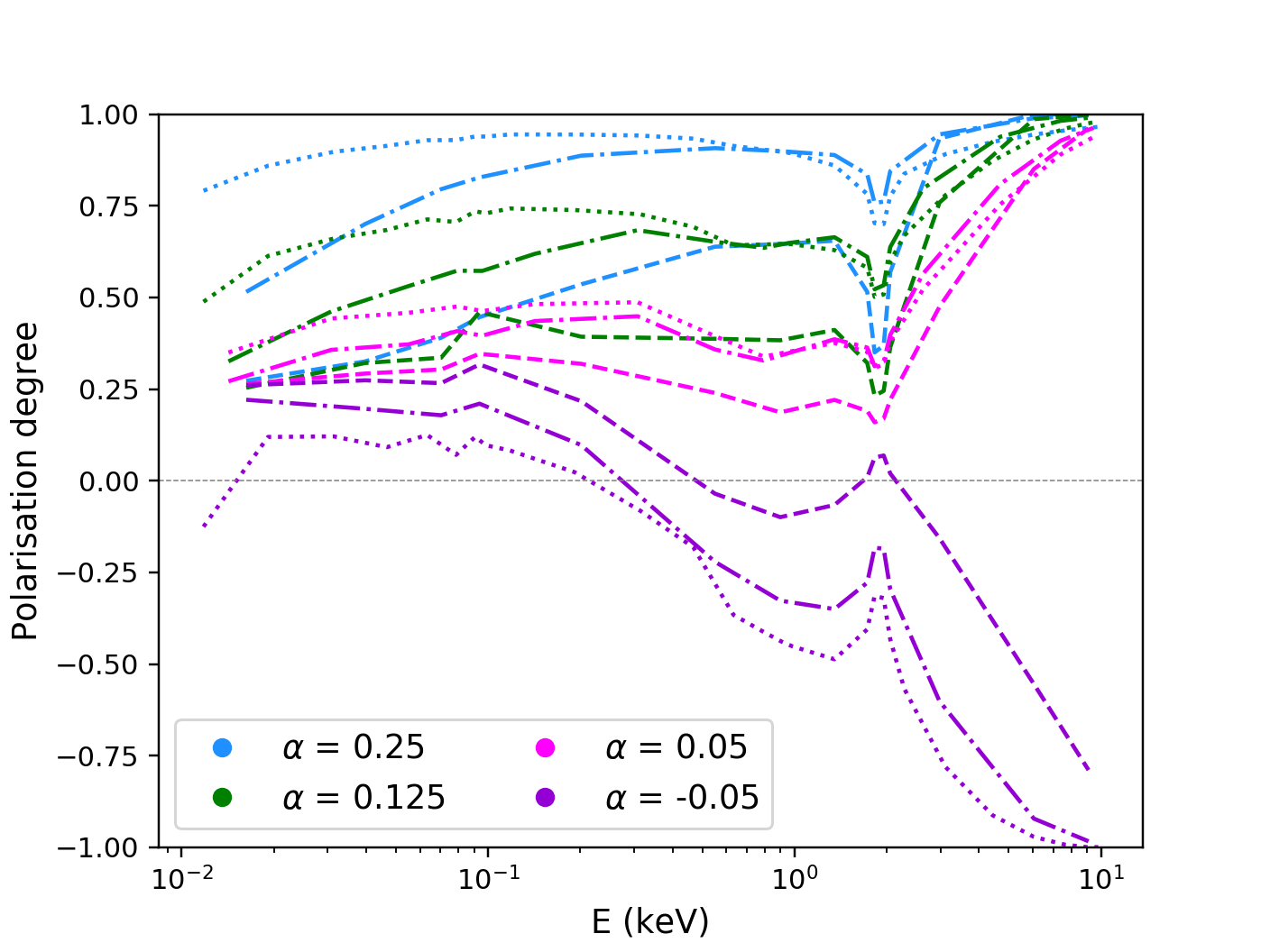}}
    \caption{Polarisation degree as a function of the photon energy $E$ for atmospheres experiencing particle bombardment with the temperature profiles shown in Figure \ref{fig:PL_temp} (line code is the same here).}
    \label{fig:PL}
\end{figure}

To explore this issue further, 
we considered the set of (ad hoc) temperature profiles discussed in Section \ref{sec:temp-prof}, computed by assuming a particle bombardment profile for $B=3\times10^{14}\ \mathrm{G}$, $L_\infty=10^{36}\ \mathrm{erg/s}$  and $y_0 = 100\ \mathrm{g/cm}^2$ (model 1 in Table \ref{tab:modelvalues}) in the external region below a given column density $y_*$, and a power-law profile, with index $\alpha$, for $y \geq y_*$ (see Figure \ref{fig:PL_temp}).  
In order to investigate the effects of an ``inverted temperature profile'', we also considered a power-law with  slightly negative slopes. 

The results of our radiative transfer calculation  are shown in Figure \ref{fig:PL} for different values of $y_*$ and $\alpha$. If the  temperature profile at some point starts increasing with depth, as expected in the presence of crustal heating, the emergent signal is always  dominated by the X-mode at all frequencies.
Models in which the temperature starts increasing at a lower depth, closer to the top of the atmosphere and to the Compton-cooled bombarded layers, have larger polarisation degrees in the lower energy range ($<1\ \mathrm{keV}$) when compared with models with the same value of $\alpha$ but larger $y_*$. 
Steeper slopes  also results in a polarisation degree which increases more rapidly with energy. However, in all cases we considered, if the internal temperature profile increases inward (positive $\alpha$) the emergent signal tends to be polarised $100\%$ in the X-mode at $10\ \mathrm{keV}$. 
In contrast, if for some reasons the 
temperature profile decreases  towards the base of the atmosphere, 
then  a change in the dominant polarisation mode occurs, with an emergent signal that  becomes O-mode dominated at high energies.
 \begin{figure*}
     \centering
     \subfigure[]{\includegraphics[width=\columnwidth]{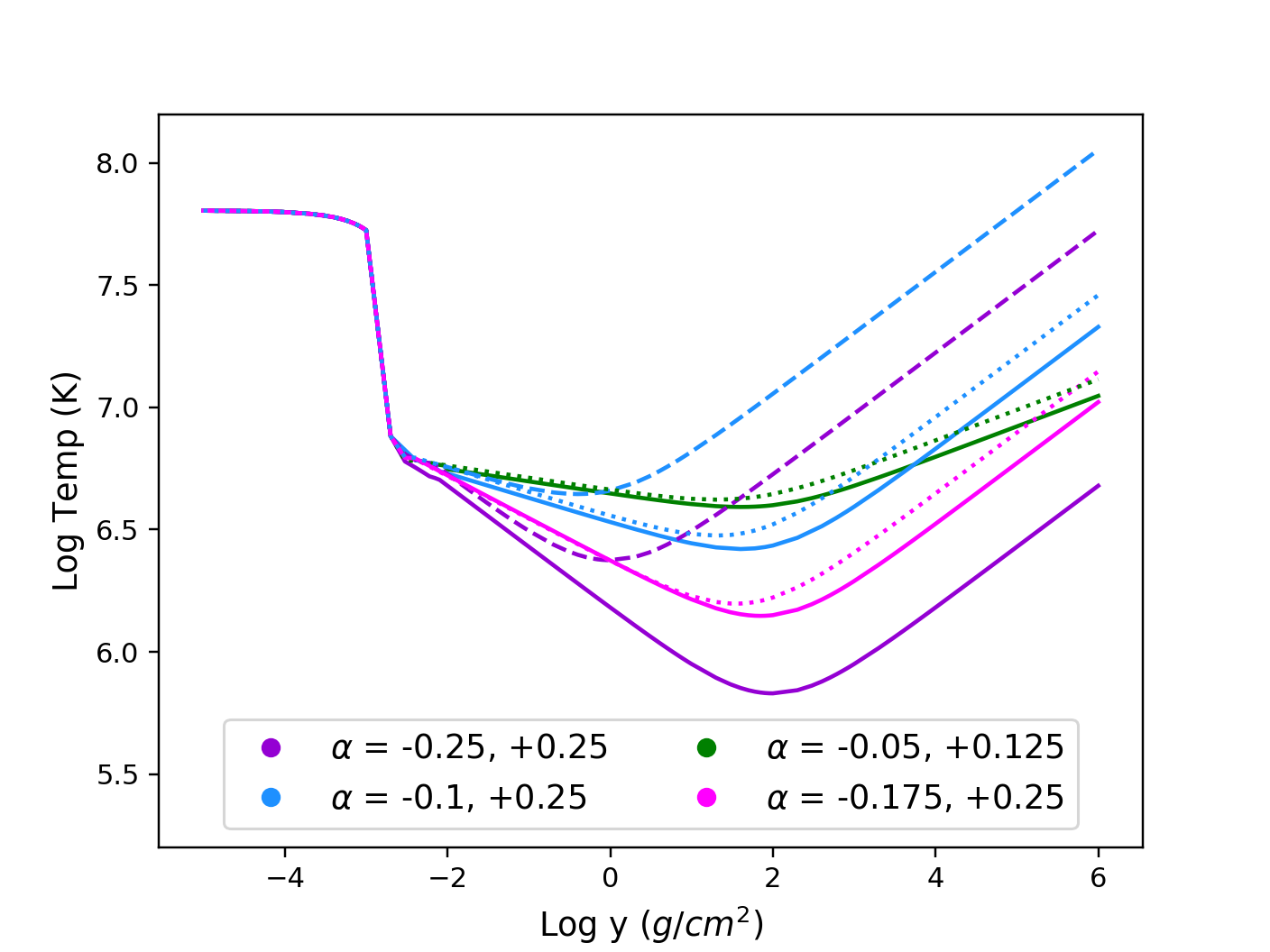}}
     \subfigure[]{\includegraphics[width=\columnwidth]{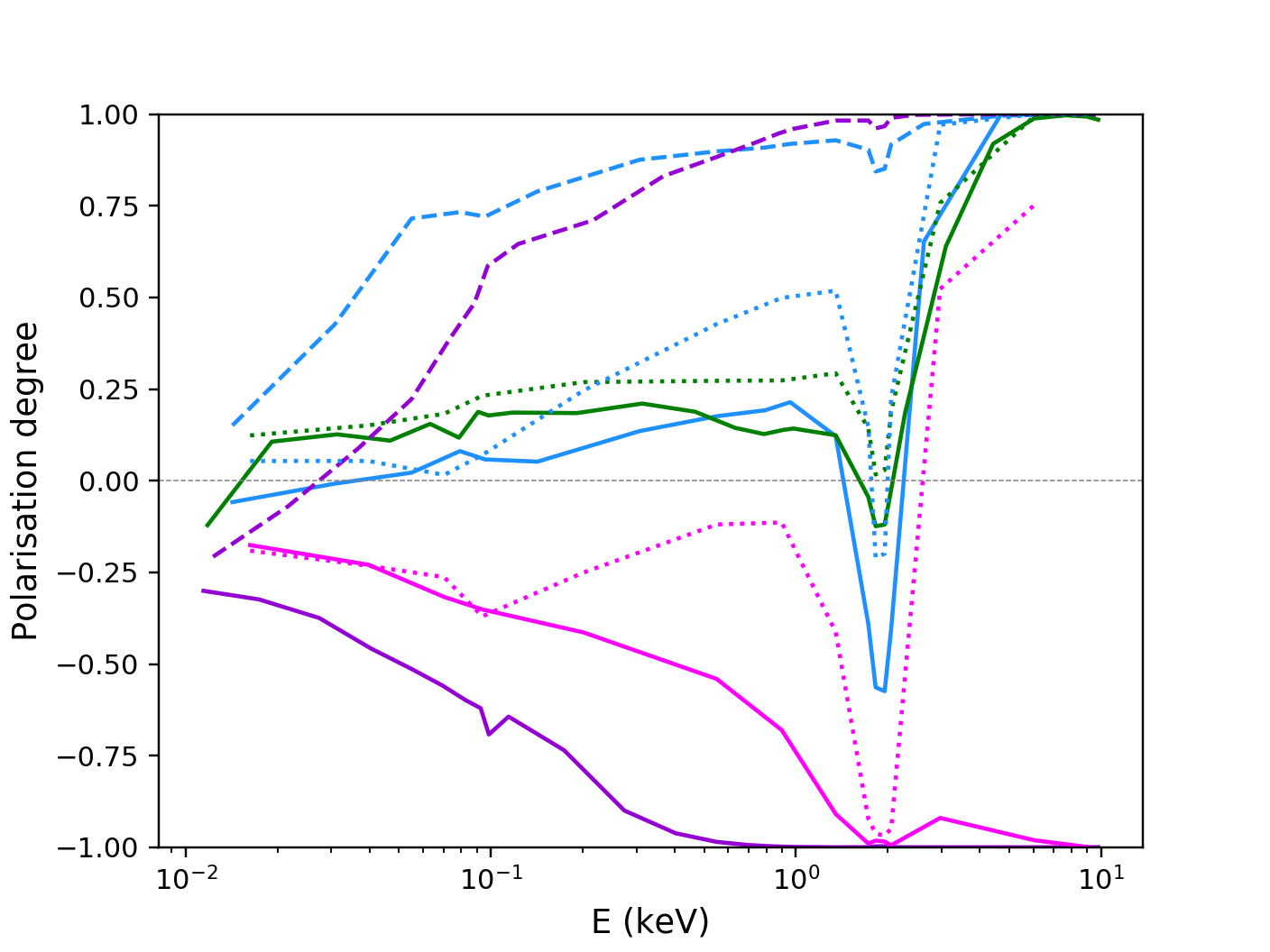}}
     \caption{Temperature (a) and polarisation degree (b) for atmospheres experiencing particle bombardment in the most external layer (same model used in Figure \ref{fig:PL_temp}), but 
     with a double power law temperature profile. The temperature is first assumed to decrease with index $\alpha_1 = -0.25, -0.175, -0.1, -0.05$, starting from a depth $\log y_1=-2$ and then to increase with index $\alpha_2 = 0.25, 0.125$ starting from a depth $\log y_2=2$ (solid lines), $\log y_2 = 1.7$ (dotted lines) and $\log y_2 =0$ (dashed lines).}
     \label{fig:neg_pos}
 \end{figure*}
 
Motivated by this, as a final test, we considered models in which 
the external layers ($y \leq 10^{-3}$~g cm$^{-2}$) are still assumed to be heated by particle bombardment (model 1 is assumed in the following), but the temperature profile follows a broken power law as $y$ increases. The first power law component decreases with depth and extends in an intermediate layer, while the  second one increases inward and is applied to the innermost regions. 
For the increasing leg we considered the reference slope $\alpha_2=1/4$ of the gray, 
passive cooling atmosphere, and a case with half this value. For the decreasing part we investigated cases with a slightly decreasing behaviour ($\alpha_1=-0.05\,,-0.1$, so to mimic a small deviation from the completely flat profiles of Figure \ref{fig:DenisTemp}) and two more extreme, albeit unphysical, cases with  $\alpha_1=-1/4\,,-1/8$. We also explored the effect of varying the extension of the two regions.
The temperature profiles  are illustrated in Figure \ref{fig:neg_pos}(a) and the results of the radiative transfer calculation are shown in Figure \ref{fig:neg_pos}(b). 

For models with $\alpha_1\le 0.1$ (blue and green curves in Figure \ref{fig:neg_pos}), the emergent signal is dominated by the X-mode in almost the entire energy range, with the exception of the very low energies below $\sim 3\times 10^{-2}\ \mathrm{keV}$ and, possibly, of the region near the proton cyclotron energy. This is also true for all models in which the temperature starts rising at a lower depth (i.e. $\log y_2 = 0$).
The polarisation degree increases with energy (apart from the depolarisation which appears near the cyclotron energy).
Higher temperatures deeper in the atmosphere result in higher polarisation degrees across the whole energy range. The proton cyclotron line is present in all spectra.
In order to produce an O-dominated signal in a 
wide energy band, a very steep inverted temperature profile, extending up to the entire stopping length (i.e. $y_2=y_0=100\, \mathrm{g/cm}^2$) is required, as shown by the purple and magenta solid lines in Figure \ref{fig:neg_pos}. 
We have only produced one scenario, when the knee in the broken temperature power-law occurs at $\log y = 1.7$ (magenta dotted curve in Figure \ref{fig:neg_pos}) where the emission  switches from O- to X-mode dominated close to the {\it IXPE} energy range ($2$--$8\ \mathrm{keV}$).
The dominance switch from O-mode to X-mode can be seen around the cyclotron energy.
However, for this switch to occur in and around the {\it IXPE} energy range a very steep inverted slope is necessary, and this is therefore an unrealistic model.

\section{Application to Sources}
\label{sec:application}

In order to compare our results with the recent observations of magnetars performed by {\it IXPE}, we need to compute the model spectro-polarimetric properties as seen by a distant observer. To this aim, we use a ray-tracing technique which sums together the contributions from the parts of the star surface which are in view at each rotational phase \cite[][see also \citealt{taverna_polarization_2015, gonzalez_caniulef_polarized_2016}]{zane_unveiling_2006}. In doing this, we first compute the monochromatic, phase-dependent flux of the three Stokes parameters $I,\,U,\,Q$ and then calculate the observed flux, polarisation degree, $\sqrt{Q^{2}+U^{2}}/I$, and angle, $\arctan(U/Q)/2$, at each rotational phase. We assume a core-centred dipole magnetic field and include general relativistic corrections.

We first address the question of whether particle bombardment can be a viable explanation for the modest polarisation observed from these sources at low energy, in alternative to condensed surface emission, as suggested by \citet{taverna_polarized_2022} and \citet{zane_strong_2023}. 
We assume two different geometries for the bombarded atmospheric patch: a polar cap with 
semi-aperture of $5^\circ$, 
and an equatorial region with semi-aperture of $5^\circ$ in both latitude and azimuth. 
This is motivated by the fact that similar geometries have been invoked to explain the polarisation at low energies  measured by {\it IXPE} in 4U 0142+61 and 1RXS J1708 \cite[][]{taverna_polarized_2022,zane_strong_2023}. 
In both cases, we divided the surface region into two patches in co-latitude, such that the magnetic field inclination in each patch is $\theta_B = (0^\circ, 5^\circ)$ and $(85^\circ, 89^\circ)$ for the cap and the equatorial spot,  respectively. Radiative transfer models have been computed by accounting for the inclination of the magnetic field, which introduces an azimuthal dependence in the radiation field. For the calculation, we used the same grids  as in the previous section, and in addition an azimuthal mesh with $5$ points 
such that the ray direction and magnetic field is calculated for each $(\mu,\,\phi)$ pair (see  \citealt{lloyd_model_2003},  Section 4.2.2, for more details). 
Here we discuss the limiting case 
in which the emission is produced only by the particle bombardment, since this is the scenario in which we expect the emergent signal to have a suitably low polarisation degree.
We used model 1 \footnote{The models have been computed in terms of observed effective temperature.  The luminosity is then correctly associated to each surface patch accounting for the corresponding patch area.} from Table \ref{tab:modelvalues}, with $B = 3\times10^{14}\ \mathrm{G}$, $L_\infty = 10^{36}\ \mathrm{erg/s}$ and $y_0 = 100\ \mathrm{g}/\mathrm{cm}^2$.
For the rest of the surface, we assumed unpolarised blackbody emission at a very low temperature such that the contribution to the total spectrum is negligible.

 \begin{figure*}
     \centering
     \subfigure[]{\includegraphics[width=\columnwidth]{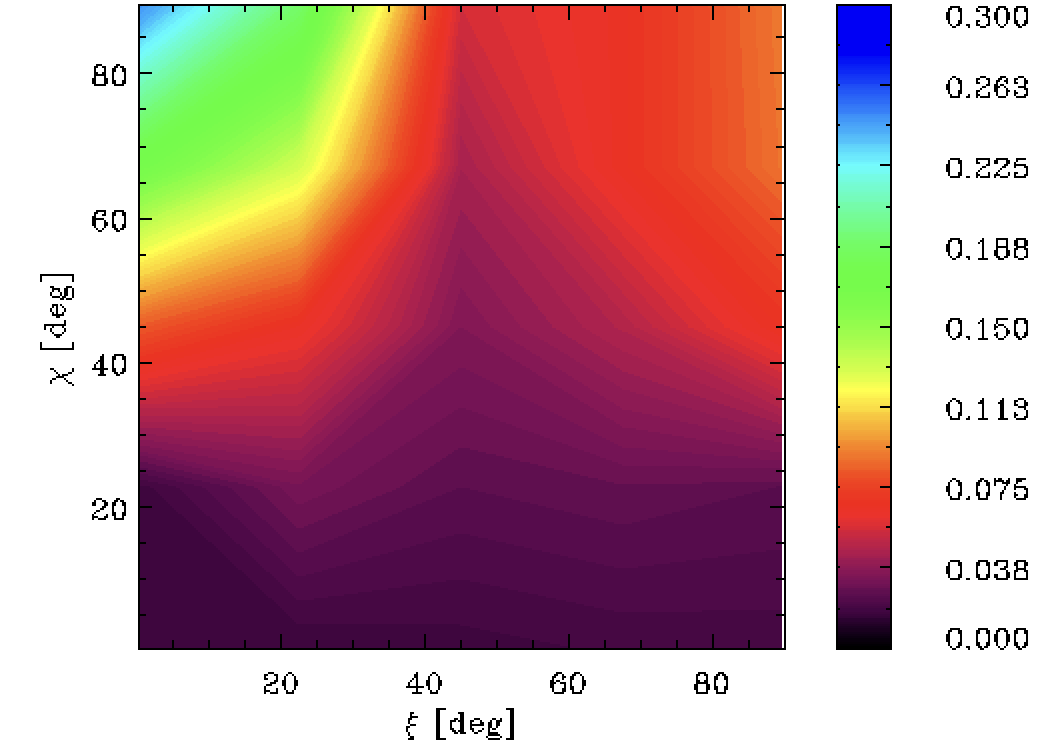}}
     \subfigure[]{\includegraphics[width=\columnwidth]{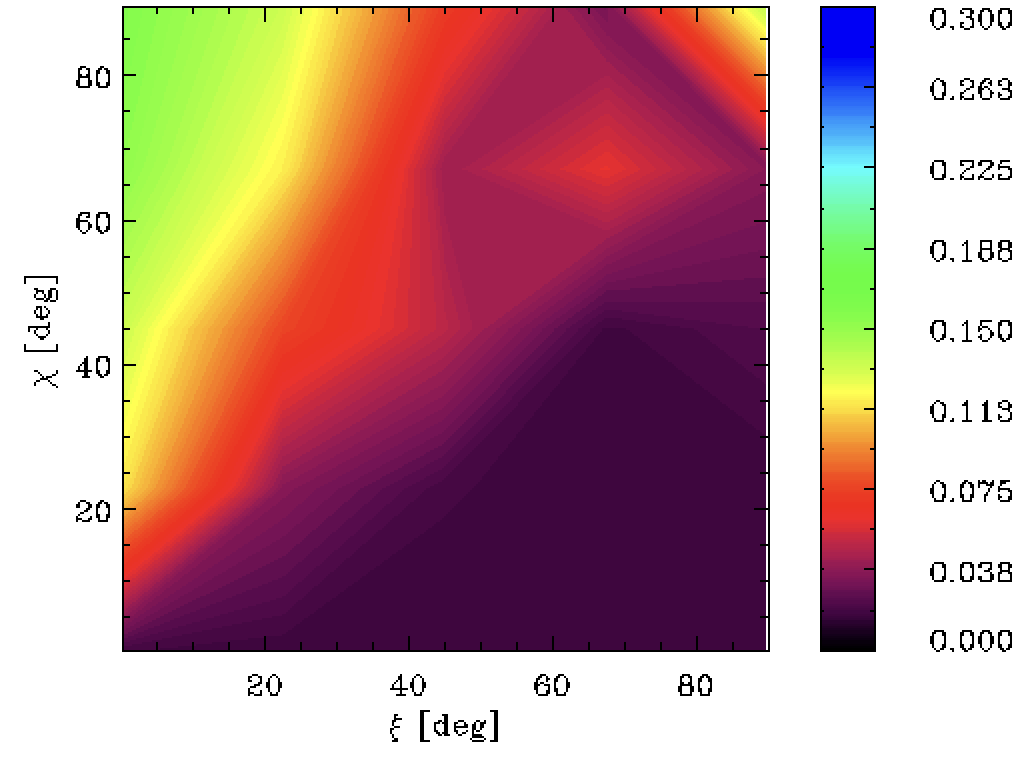}}
     \caption{The phase-averaged polarisation degree in the $2$--$3\ \mathrm{keV}$ energy range as function of the two angles $\chi$ and $\xi$ for a bombarded atmosphere on top a polar cap (a) and an equatorial patch (b). See text for model details.}
     \label{fig:ray}
 \end{figure*}

Figure \ref{fig:ray} shows the phase-averaged polarisation degree,  in the $2$--$3\ \mathrm{keV}$  energy range, as seen by a distant observer, as a function of the angles that the spin axis makes with the line-of-sight, $\chi$, and the magnetic axis, $\xi$, respectively.
We find that the maximum observed polarisation degree is $\sim25\%$ in the polar cap model (Figure~\ref{fig:ray}a), while it is slightly lower for the equatorial spot (Figure~\ref{fig:ray}b). Both of these are 
compatible with the polarisation properties observed from 1RXS J1708 in the same energy range.
Emission from a hot spot heated by particle bombardment could therefore explain {\it IXPE} measurement at low energies. 

Additionally, the polarisation degree observed from 4U 0142+61 could potentially agree with our models. The $90^{\circ}$ rotation in polarisation angle between the low and high energy ranges cannot be accommodated in this scenario alone. However, including RCS \cite[as proposed by][]{taverna_polarized_2022}, particle bombardment certainly cannot be ruled out as an explanation for this magnetar either. We have not produced models of the emission at infinity for the case from Figure \ref{fig:neg_pos} in which the polarisation goes from O-mode dominated to X-mode dominated at $\sim 3\ \mathrm{keV}$ (magenta dotted curve) because for this to occur an unrealistically, unphysically large decrease in temperature would be needed and we have no astrophysical reason to believe that this would ever occur.

\section{Discussion and Conclusions}
\label{sec:discussion}

In this work we have investigated the polarisation properties from magnetar atmospheres experiencing particle bombardment. We computed simplified models 
using ad-hoc assumptions on the temperature profile, mimicking scenarios in which the emission is powered by different combinations of particle bombardment and heat injection from the crust.  We also considered extreme cases in which some internal atmospheric layers are characterised by a smooth, inverted thermal profile that slightly decreases inward. 
We have found that particle bombardment can, in principle, produce a distinct polarisation signature.

When assuming that the luminosity is only due to the release of heat from particle bombardment, with no significant contribution from the crustal heating, we find that the magnetar X-ray emission has a polarisation degree that remains around $20\%$ below $\approx 0.2\ \mathrm{keV}$, dips slightly and then increases $\approx 5$--$40\%$ between $1$--$10\ \mathrm{keV}$. The emission is X-mode dominated.

Unsurprisingly, when the contribution to the total luminosity from crustal heating is increased, polarisation properties tend towards those of a standard  cooling atmosphere. Atmospheres that exhibit a steeper increase in temperature in the internal layer produce radiation with higher polarisation degrees, while moving inward the depth at which the temperature starts to  increase results in a less polarised signal.
Additionally, inverted profiles, in which the temperature decreases with depth, may in principle result in O-mode dominated emission in the X-ray band, although the required slope appears to be unphysically large.
However, in the case of particle bombardment onto a non-magnetised atmosphere, it has been been found that particles with low Lorentz factor ($\gamma \sim 10$) bombarding the atmosphere can produce heat deposition in a very shallow layer and give raise to an inner region with a relatively steep inverted temperature slope \cite[]{baubock_atmospheric_2019}. This study is not directly applicable to the regime we are using in this work, among other reasons because in our study we account for the fact that the length within which heat is deposited is larger than the stopping length of primary charges (see section \ref{sec:temp-prof}). An in-depth study and proper calculation therefore need to be carried out to investigate if a similar scenario could be possible in the magnetised plasma of a magnetar atmosphere and if these seemingly unphysical temperature slopes, required for an O-mode dominated spectrum, can in fact be produced.

Particle bombardment alone cannot fully explain the polarisation features observed by {\it IXPE} in magnetar sources. However, a bombarded atmosphere combined with other emission models (like a standard atmosphere, RCS, a condensed surface, etc.) may provide a viable explanation for the polarisation pattern detected at low energies. Specifically, in the case of 1RXS J1708, particle bombardment is in agreement with the emission in the $2$--$3$ keV range. We speculate that a hot spot heated by particle bombardment coupled with standard atmospheric emission would be a likely scenario, as proposed by \cite{zane_strong_2023}, as an alternative to the one based on a combination of gaseous and condensed components. Additionally, models that combine particle bombardment and magnetospheric RCS could also potentially explain the emission from both 4U 0142+61 and SGR 1806–20. Testing these combined models is outside the scope of this current paper but is an exciting prospect for future work.

It is predicted that the external magnetic fields of magnetars are complex and contain local twists, with particle bombardment taking place only in a small selection of magnetic colatitudes. However, the implementation of a more complex field structure is not achievable for this work. Instead, we produce the emission at infinity assuming a global dipolar field in our ray-tracing simulation and allow contributions to the emission spectrum from only small patches of the surface. While this simplification of the magnetic structure will likely have an impact on the polarisation spectrum, the global twist is a well-known and commonly used approximation \cite[see e.g.][]{thompson_electrodynamics_2002, fernandez_x-ray_2011, taverna_probing_2014}. The development of a model including complex magnetic field structures and the investigation of the magnetic field evolution are important areas of ongoing and future works.

In principle, to fully probe the effects of particle bombardment the full thermal and pressure structure of the atmosphere should be solved coupled with radiative transfer, relaxing the assumptions of radiative and hydrostatic equilibrium. Additionally, our models assume a fully-ionised hydrogen plasma and do not include vacuum contributions to the plasma dielectric tensor. Vacuum birefringence and mode conversion effects can significantly affect the polarisation properties of the emission around and above the quantum critical field \cite[]{zane_magnetized_2000, ozel_surface_2001, ho_ii_2003, ho_iii_2003, lai_transfer_2003, kelly_x-ray_2024}. The focus of this paper was to explore the potential polarisation signatures from particle bombardment on the atmosphere and further developments will be matter of future investigations.

\section*{Acknowledgements}

RK acknowledges support from The Science and Technology Facilities Council (STFC) for a PhD studentship.
RTa and RTu acknowledge financial support from the Italian MUR through the grant PRIN 2022LWPEXW.
DGC acknowledges support from a CNES fellowship. We would like to thank the anonymous referee for their insightful comments and helpful suggestions.
\section*{Data Availability}
The simulated data produced in this investigation are available on request.



\bibliographystyle{mnras}
\bibliography{references} 







\bsp	
\label{lastpage}
\end{document}